# Prosocial Design in Trust and Safety

David Grüning[1, 2, 3]*

Julia Kamin[1, 4]*

1: Prosocial Design Network

2: Max-Planck Institute for Human Development, Center for Adaptive Rationality

3: Stanford University, Stanford Mussallem Center for Biodesign

4: Civic Health Project

*Both are corresponding authors: david@prosocialdesign.org, julia@prosocialdesign.org.

**Abstract**

This chapter presents an overview of Prosocial Design, an approach to platform design and governance that recognizes design choices influence behavior and that those choices can–or should– be made toward supporting healthy interactions and other prosocial outcomes. The authors discuss several core principles of Prosocial Design and its relationship to Trust and Safety and other related fields. As a primary contribution, the chapter reviews relevant research to demonstrate how Prosocial Design can be an effective approach to reducing rule-breaking and other harmful behavior and how it can help to stem the spread of harmful misinformation. Prosocial Design is a nascent and evolving field and research is still limited. The authors hope this chapter will not only inspire more research and the adoption of a prosocial design approach, but that it will also provoke discussion about the principles of Prosocial Design and its potential to support Trust and Safety.

**Introduction**

Prosocial Design is an approach to platform design and governance that, at its root, acknowledges that design choices influence the behavior of users and those choices can–or should–be made toward producing prosocial outcomes. Some of its goals are shared with Trust and Safety, such as ensuring users feel safe to engage on a platform, yet the aims of Prosocial Design can expand beyond the scope of Trust and Safety's mandate to, for example, foster understanding and empower communities. A second distinguishing factor of Prosocial Design is that it focuses less on identifying and stopping malicious ("bad faith") actors and content, and instead centers on creating the conditions for "good faith" users to engage in less harmful ways and to participate in healthier interactions.

In this chapter, we first offer several principles of Prosocial Design, discuss where there are tensions within those principles, and describe where Trust and Safety and Prosocial Design align and diverge. Next, and as the main contribution of this chapter, we provide a brief review of prosocial design solutions as they apply to two goals shared with Trust and Safety, minimizing harmful behavior and reducing misinformation, as well as offer applications to adjacent goals such as user wellbeing and healthy discourse. Prosocial Design is a nascent field whose scope and principles have yet to solidify and where research is still limited. We are hopeful this chapter will not only inspire more research and wider adoption of Prosocial Design, but that it will also spark discussion about what are–and are not–its core elements.

**Defining Prosocial Design and its principles**

As with many fields, particularly new ones, there is no canonical definition of "Prosocial Design." The two authors of this chapter lead the Science Board of the Prosocial Design Network, a network of over 400 researchers, technologists and civic society leaders. Drawing from the insights of our members and for the purposes of giving them a common framework to build from and, we have developed the following working definition of Prosocial Design, which we also use for this chapter:

> "Prosocial Design is the set of design patterns, features and processes which foster healthy interactions between individuals and which create the conditions for those interactions to thrive by ensuring individuals' safety, wellbeing and dignity."

This definition captures both how the term "prosocial design" is most often conceptualized and also indicates one of its core principles - i.e. it is design that *leads to* prosocial behavior (Oliveira et al., 2021) by "encouraging, affording, and amplifying" those behaviors (Schirch et al, 2023). But the definition above also highlights what many in the Prosocial Design Network see as essential foundations for prosocial design; that it first needs to ensure individuals on a platform are safe, their wellbeing is not being harmed and their dignity is honored. Indeed, it would not be possible to foster healthy social outcomes without a foundation of safety, wellbeing and dignity.

Prosocial Design, again, is an evolving field that includes researchers, technologists and advocates who each have a unique understanding of what is its essence, yet there are several principles we see commonly shared. Those principles include:

- *Dignity of the user*: Prosocial Design views users as humans who deserve dignity. This dignity is afforded even to users whose behavior may be harmful and breaks platform rules; in those cases, Prosocial Design might integrate procedural justice practices by, for example, giving users an explanation for moderation decisions and a means to appeal that decision. Centering the dignity of the user also encourages a tendency toward designs that allow for - or even strengthen - user agency, giving users the reins to make choices they feel best. It also encourages - if not requires - transparency, avoiding deceptive practices even if they may be considered to benefit other users.

- *Design is not neutral:* A core premise of Prosocial Design is that design choices have an impact on user behavior; those choices can inadvertently lead to harmful behaviors or can be more intentionally selected to produce more prosocial interactions.
- *More proactive than reactive:* As a possible corollary to the previous principle, Prosocial Design has a bias toward design choices that are more proactive - or upstream - that can foster prosocial behavior, as opposed to reactive or downstream design that respond to antisocial behavior. New_ Public (2022) uses the analogy of designing public spaces which is useful in thinking about proactive design: the design choices made for a space - whether physical or digital - signal to people how to engage and set the tone for what happens in those spaces.

It may occur to the reader that the above principles can sometimes be in tension with each other. In particular, a critic could argue that any design meant to influence (or "nudge") behavior is inherently manipulative and paternalistic, so does not afford users full agency and dignity (Schmidt & Engelen, 2020). Profit oriented platforms are known to often use 'deceptive patterns' that manipulate users toward the goal of increasing revenue (Brignull et al., 2025); employing design patterns with prosocial goals may be beneficial to society or even the individual user but is no less manipulative, and is paternalistic in that it assumes the designer knows what is best for individual users (Hertwig & Grüne-Yanoff, 2017). The premise that design can create prosocial outcomes itself risks being "tech solutionist". (Badiei & Fidler, 2021)

This line of criticism also represents a healthy tension in the views and conversations among those within the Prosocial Design community. Some prioritize design that supports individual agency and rights as prosocial. Others take the view that since there is no such thing as neutral design, designers must take responsibility for how their choices result in either anti-social or prosocial outcomes; giving users agency to decide which design choices are best is not a realistic solution as it can be a burden on users, most of whom may not have the bandwidth or desire to exercise their agency. We suggest there is a healthy and viable middle ground between these two views of Prosocial Design; one that acknowledges that design should be oriented toward prosocial outcomes, but that it also should strive to support user agency and dignity by, for example, being transparent about both its operationalization and goals, affording ways to "opt out" when possible (Hertwig & Grüne-Yanoff, 2017) and, when appropriate, by "boosting" user competencies to navigate their online experience (e.g., Grüning et al., 2023; Haliburton et al., 2024).

Yet another area of where there is not uniformity of thought is in itemizing the set of goals that Prosocial Design should aim for and deciding where to strike a balance when two goals are in tension. In our definition of Prosocial Design above we lay out the goals of healthy interactions, safety and wellbeing (all while respecting user dignity), but within those broad goals there is ample room for interpretation and disagreement. For example, "healthy interaction" could be discussions that prioritize empathy and ensure inclusion or could instead be vigorous debates that prize fact-based reasoning. Somewhat akin to the tension between "Safety" and "Expression" in Trust and Safety (Hunsberger, 2024), it is not possible for Prosocial Design choices to equally weight and integrate all prosocial goals simultaneously. Platform designers, community stewards and others who make design choices for digital spaces will need to prioritize the outcomes they value most, taking into account the unique purpose and audience of their online space, and weigh design choices accordingly.

**Prosocial Design in relationship to Trust and Safety and other related frameworks**

Prosocial Design and Trust and Safety overlap in several of their goals as well as approaches. TSPA's T&S Curriculum defines Trust and Safety as "fundamentally … about enabling users to have the best experiences possible" and "creating an environment or experience that proactively and affirmatively helps users achieve their objectives" (Trust and Safety Professional

Association, 2025). It also sees "preventing, detecting, and responding to abuse" as a "critical component" of Trust and Safety. Prosocial Design shares Trust and Safety's approach of putting users and their goals at the center of design. The distinction between the two fields is less a matter of differing principles and more about where their work is focused. Whereas Trust and Safety sets a primary goal of minimizing abuse, for example, Prosocial Design imagines how design can foster positive interactions and experiences. Also, while both aim to minimize harm, Trust and Safety engages both in countering "bad" actors who are intent on causing harm as well as minimizing harmful behavior on the part of average–but erring–users. Prosocial Design, in contrast, is built primarily with that second set of everyday users in mind, understanding even the best of us can, for example, unwittingly spread disinformation or use harmful language when emotions are triggered.

Prosocial Design similarly overlaps with and draws from countless other frameworks and efforts all aimed toward creating healthier online experiences. New_Public's Civic Signals, a set of goals that healthy online spaces should aspire to that was developed with experts and validated with public surveys across twenty countries, provides an organizing framework for the goals Prosocial Design can aim to achieve (2022). The theoretical (e.g., Tyler, 2016) and applied work (e.g., Katsaros et al., 2024) in Procedural Justice theory, Trauma-Informed Design (Randazzo et al., 2023) and Boosting theory (Hertwig & Grüne-Yanoff, 2017; Herzog & Hertwig, 2025) inform the principles of Prosocial Design and do much of the heavy lifting in conceptualizing how to balance safety and dignity. And just as Prosocial Design is a large tent that absorbs related frameworks, it is also an approach that is integrated into other and larger frameworks for example Safety by Design (Trust and Safety Professional Association, 2025) and the Council on Technology and Social Cohesion's Blueprint for Prosocial Tech Design Governance (Schirch, 2025).

**Selective review of Prosocial Design research**

To illustrate the potential of Prosocial Design to create healthy online experiences, we present this selective review of prosocial design patterns. We focus on two areas most closely aligned with the interests of Trust and Safety, reducing harmful behavior and mitigating misinformation, but also present use cases for goals that may expand beyond the immediate interests of Trust and Safety, for example fostering healthy discourse. The design patterns we include are drawn from Prosocial Design Network's library of evidence-based design solutions. This library does not contain all existing prosocial design patterns, let alone all imaginable patterns; it is limited to designs that have been tested and for which there is public evidence that they are effective in producing prosocial outcomes. In most cases evidence includes statistical and inferential models that lets us say with some degree of confidence the design leads to prosocial outcomes.

In the review below, we use a framework familiar to practitioners and recently introduced among researchers (Grüning et al., 2024) which categorizes designs relative to the point in time they intervene with user engagement. Specifically, it distinguishes between proactive, interactive and reactive designs. Proactive designs aim to prevent harmful behavior and encourage healthy interactions by being placed upstream to user engagements such as sharing news or responding to a post. (e.g., Bor et al., 2020; Stroud et al., 2016). Interactive interventions are implemented at the point of engagement, for example when a user is typing a comment or reading a post with a news link (e.g. Argyle et al., 2023; Renault et al., 2023). Reactive interventions are designed to reduce the likelihood that such behavior recurs (e.g., Horta Ribeiro et al., 2023; Srinivasan et al., 2019; Jhaver et al., 2019).

**Prosocial Designs to reduce harmful and other rule-breaking behavior**

Of top concern to Trust and Safety teams will be designs that reduce behavior that could harm other users, both that explicitly break platform rules but also may fall just short of platform violations. Prosocial design researchers have likewise focused their attention on patterns that can increase rule-adherence among users, including among previous rule-breakers. In all cases, researchers assume a subset of potentially rule-breaking users are not solely intent on causing harm but that to some extent they are acting in good faith and are either unaware of the rules and norms or otherwise capable of reform.

**Proactive examples.** One of the most widely studied - and now adopted - uses of Prosocial Design is to explicitly inform or remind users of the norms of an online space. At least three field experiments - conducted on Reddit, Nextdoor and Facebook - demonstrate the effectiveness in reducing rule-breaking and other harmful behavior by notifying users of what behavior is expected and what is not permitted. Working with moderators from r/Science, Matias (2019) observed that using a "sticky note" that reminds users of a subreddit's rules both increases the likelihood a given post followed those rules and also increases new users' willingness to post. Kim et al. (2022) similarly found that posting norms as users entered a group forum reduced the number of comments reported for abuse. Finally Tyler et al. (2021) looked at the impact of reminding Facebook users of the rules after they had been suspended and observed an increased adherence to rules.

While the research demonstrating the effectiveness of posting norms is strong, there is less certainty about what is the precise set of mechanisms that produces improved behavior. At least to some degree, we can expect that signaling norms is effective because people by and large want to behave as expected and only transgress when they misunderstand the norms (Vriens & Andrighetto, 2025). But as Matias (2019) and Tyler et al. (2021) caution, rules reminders may only be effective if they are likewise clear about the consequences of infractions.

**Interactive examples.** Several design solutions intervene at the point when users are crafting content to similarly remind them of platform norms or otherwise give them an opportunity to pause and reflect on their actions (Katsaros et al., 2022; Horta Ribeiro et al., 2025). Two field experiments, one conducted by OpenWeb, a service that provides comments sections (Goldberg et al., 2020), and one conducted on Twitter (Katsaros et al., 2021), test the use of an interstitial that pops up when a user has included harmful language in a comment, giving the user an opportunity to rewrite their comment before posting. Both studies see a substantial change in commenting behavior, with Goldberg (2020) noting a 12.5% increase in civil comments and Katsaros et al. (2021) observing 6% fewer offensive tweets. Chang et al. (2022) take a similar approach giving users on a subreddit in time feedback on whether their comment is likely to increase (or decrease) tensions in a thread. Yet another interstitial tool, one tested in collaboration with Reddit (Horta Ribeiro et al., 2024) and offered to subreddits, gives mods the ability to customize feedback to users when they are breaking a specific subreddit rule.

As with proactive norms reminders, there are likely multiple mechanisms at play to explain why these interstitials and in time feedback tools are effective. Like with norms reminders, intersitials may simply inform users of rules they may have been unaware of and give users a chance to course correct. Researchers, however, suggest that inserting "friction" into the process gives users a moment to pause and reflect on their choice of language, shifting them from an automatic reaction to a more thoughtful thinking that can result in a less heated response.

**Reactive examples.** The design solutions above aim to reduce the likelihood that users will produce harmful content, but of course many will choose to ignore reminders and nudges and post abusive or otherwise rule-breaking content. Research shows, however, there is still an opportunity to engage with rule-breaking users in a way to produce prosocial outcomes. Jhaver et al. (2019) looked at millions of posts across hundreds of subreddits and found that giving users an explanation for why their post was removed reduced the user's likelihood of having a future post removed. The effect was stronger for longer explanations, and equally effective when delivered by a mod or a bot.

There is also evidence that the removal of harmful content itself can have a prosocial outcome on the rule-breaker. Ribeiro et al. (2022) and Srinivasan et al. (2019) use quasi-experiments with data from Facebook and Reddit and observe that removing harmful content reduces recidivism. It is important to note that in both cases, users were given information about an appeals process and future consequences and, in the case of Reddit users, were told why their content was removed, so it may be the case that content removal is only effective when accompanied by such due process steps. This interpretation is further supported by two field experiments on Nextdoor (Katsaros et al., 2024; Katsaros & Grüning, in preparation), in which the researchers find that hiding offensive comments, as opposed to removing them and notifying users, had no significant effects on user behavior.

**Prosocial Designs to reduce the spread of misinformation**

A second area of prosocial design research that may be relevant to Trust and Safety and Integrity teams are interventions that reduce the spread of misinformation. While stemming misinformation may be of secondary interest to some Trust and Safety teams - compared to, for example, teen mental health or identifying malevolent actors - we include it in this review because it is the focus of the majority of academic research related to Prosocial Design. That may be the case, in part, because it lends itself to Prosocial Design; because of the challenges of precisely identifying misinformation and of concerns over censorship, prosocial designs that support users in discerning what is true or false have become an important lever in curbing the spread of misinformation. We see reducing misinformation as a prosocial goal and as relevant to Trust and Safety both because misinformation can harm individuals, for example by giving poor health advice, but equally or more so because of the potential societal impacts, like dangerous social division or swayed elections.

**Proactive examples.** Researchers have investigated several strategies to proactively arm users with the ability to discern true from false information. Two approaches, which are often both referred to as "inoculation," aim to give users the skills and information they need to recognize misinformation when they come across it. One of those approaches, which we term "misinformation literacy," inoculates by training users to recognize the strategies purveyors of misinformation tend to use, such as emotional language or false dichotomies, across any topic or issue. This approach has been tested using interactive games and engaging video tutorials (90 seconds to 15 minutes) demonstrating their potential effectiveness (e.g., Bor et al., 2020; Roozenbeek & van der Linden, 2019; Roozenbeek et al., 2022). The second inoculation approach, which we term "prebunking", equips users with specific fact-checked information on topics that are the subject of misinformation campaigns. Three field studies (Bowles et al., 2023; Pereira et al., 2023; Yadav & Garg, 2023), which each make use of fact-checking subscriptions (delivered via text, Whatsapp or a newspaper subscription) show that prebunking can make users less susceptible to misinformation.

While both misinformation literacy and prebunking can be effective approaches, they are relatively high intensity interventions that require user buy in. A third approach, "accuracy prompts," has likewise demonstrated effectiveness but is a lighter design feature. Accuracy nudges can take many forms, such as a poll asking users about the importance of accuracy, or a message that most people value accuracy. They can also be a quick shot of "misinformation literacy" placed, for example, as a side banner with "tips" for detecting fake news. Several survey experiments, combined with a field study conducted on Facebook and Twitter, provide convincing evidence that accuracy prompts are an effective strategy to decrease the spread of misinformation (Lin et al., 2024; Pennycook & Rand, 2022).

**Interactive examples.** There are likewise several interactive approaches to decrease the impact of misinformation. Perhaps the most extensively researched are fact-checking labels, which appear alongside a news item to indicate its veracity, most often as assessed by professional fact-checkers. In several survey experiments (e.g., Clayton et al., 2020; Mena, 2019; Porter & Wood, 2022), researchers show that fact-checking labels reduce the likelihood of sharing (but see also Offer-Westort et al., 2024). There is evidence, however that these labels can backfire, for example by increasing general skepticism and, hence, overall platform engagement (Hoes et al., 2023; Hameleers & van der Meer, 2023) or by giving the misleading impression that news items that do *not* have a fact-checked label are necessarily true, known as the "implied truth effect" (Pennycook et al., 2020).

A similar approach uses labels to inform users about the credibility of the source of news content. Three studies show promising but limited evidence for the effectiveness of source labels. In two online experiments (Arnold et al., 2021; Celadin et al., 2023), researchers show that source credibility labels reduced participants' intentions to share misinformation on social media. In a field experiment (Aslett et al., 2022) participants in a treatment condition used the browser extension "NewsGuard" (www.newsguardtech.com), which rates the reliability of news sources, for several weeks; although some participants slightly reduced their consumption of low-quality news, overall the researchers did not observe a change in participants engagement with unreliable news sources when they used Newsguard.

A third interactive approach is to use "community notes", crowdsourced information that gives context to a post and that is only displayed if the note is rated as helpful by users with different views. In three quasi-experimental studies (Chuai et al., 2023; 2024; Renault et al., 2023), researchers show that once a community note is displayed, on average it will reduce reshares of a misleading post by over 50%. This dramatic effect is supported by an internal Twitter study (Wojcik et al., 2022), which reports a 25% to 34% reduction in retweets when a note is attached to a misleading post. The overall effectiveness of Community Notes drops significantly, however, when considering that, by virtue of depending on crowdsourcing, notes usually are displayed after a post received most of its reshares and that only a fraction of misleading posts ever receive a note.

**Reactive examples.** To the best of our knowledge, there exist no tested reactive interventions against misinformation online. This may be the case because early research demonstrated that debunking misinformation after someone has been exposed to it is challenging, and may even have a backfire effect (Nyhan & Riefler, 2010). Yet we imagine there may still be potential in debunking interventions that aim to counter misinformation by presenting fact-based corrections, ideally with clear explanations of why the misinformation is false and how the correct information can be verified (Lewandowsky et al., 2012). These interventions can be delivered after someone has interacted with or shared misleading content, either via notifications, fact-check labels, or personalized feedback. Although not yet tested in real-world, platform-integrated settings, reactive debunking could be effective in reducing belief in false information, particularly when corrections are timely, credible, and detailed (Chan et al., 2017; Walter & Tukachinsky, 2019).

**Prosocial Designs to achieve other goals**

Beyond mitigating harmful behavior and reducing misinformation online, Prosocial Design addresses a wide range of other goals, the most prominent of which are to enhance users' agency, increase their long-term well-being when using the platform, promote pluralism and stimulate discourse with high-quality content, and build a thriving online community. We share several illustrative examples.

**Agency and wellbeing.** Lukoff et al. (2023) developed and tested a toggle feature that lets users switch between an explore mode, which generates suggested content, and a focus mode, that presents only search results, while using a YouTube-like platform. Study results show that being able to use a toggle (compared to being in one mode constantly) gave users a greater sense of agency while on the platform as well as more overall satisfaction with that time spent. Another design aimed at decreasing distraction and improving users' satisfaction with their social media engagement is to batch notifications. Fritz et al. (2019) showed that grouping smartphone notifications into scheduled batches, delivered at specific intervals, increased participants' feelings of productivity and control over their smartphone use. Participants who received batched notifications, compared to those receiving an unpredictable stream of notifications, also reported more positive moods and less stress associated with being online. A related study conducted at Facebook (Weijun et al., 2022) provides evidence that sharing fewer, higher quality, notifications can increase engagement overtime.

**Discourse and pluralism.** Researchers have tested several designs aimed at improving the quality of discourse in online spaces. One approach, tested by Argyle et al. (2023) and which has become increasingly viable with advanced Large Language Models (LLMs), gives users suggestions on how to rephrase comments to, for example, communicate greater empathy or that another user has been heard. Similar to the interstitials discussed above, Chang et al. (2022) shows the effectiveness of giving users in time feedback on whether their comment can diffuse tensions in a thread. Healthier interactions may also be promoted by nudging users engaged in a heated thread to form a group where they can continue the discussion (Kim et al., 2022). More generally, the quality of user comments has been shown to be improved by the use of upvotes (Papakyriakopoulos et al., 2023) and by highlighting high-quality user comments (Wang & Diakopoulos, 2022). Researchers have also tested ways in which design can more specifically foster understanding and cohesion across individuals and groups with different political views. Van Loon et al. (2024) randomly assigned participants to engage in a simulated social media environment that either awarded users with an 'open-minded' badge, signaling intellectual humility, or a popularity badge, and observed that participants in the 'open-minded' badge condition produced more intellectually humble content and reported having a more positive overall experience. Additionally, studies generated from the Center for Media Engagement have demonstrated initial evidence that the use of a "Respect" button and memes that remind users of our shared humanity can increase engagement across partisan posts and decrease negative counter-partisan sentiment (Stroud et al., 2016).

**Belonging and community.** Finally, to help build a thriving community, Matias et al. in 2020 tested a "thank you" button, a feature that allows Wikipedia editors to express their gratitude for other editors' contributions. The researchers showed that among Arabic, German, Polish and Persian Wikipedia contributors, receiving a thank you increased an editor's two-week retention rate by two percentage points and, as a prosocial contagion effect, motivated people to express their gratitude to others significantly more often.

**Discussion**

In this chapter, we presented an overview of Prosocial Design, an approach that acknowledges the inherent capacity of digital design to influence behavior and proposes that platforms use that capacity responsibly to design for prosocial outcomes, while also supporting the wellbeing, safety and dignity of users. We discuss some of the central principles of Prosocial Design–dignity, non-neutral design, and proaction–and how Prosocial Design draws from and contributes to related frameworks in Responsible Tech. As a main contribution, we review examples of Prosocial Design drawn from PDN's library of evidence-based design patterns (Prosocial Design Network).

While the review above is not exhaustive of all relevant research, it does represent a good sum of the existing public research that uses experiments or other large-N inferential analyses to test the effectiveness of prosocial interventions. Which is to say, there are still few interventions that have been thoroughly tested to have a positive impact on desired prosocial outcomes. Of that existing research, moreover, the preponderance is focused on interventions that reduce the spread and impact of misinformation or that, to a lesser degree, reduce the incidence of harmful behavior.

In other words, most of the existing tested "prosocial" designs are, in fact, designs focused on reducing "antisocial" behavior. We know comparatively little about the effectiveness of designs that foster healthier interactions and that, for example, encourage curiosity, empathy or connection.

Yet we see glimmers that, to quote a recent article, it may be "prosocial design's time to shine" (Hunsberger, 2025); there is a growing movement of researchers and civil society organizations that are advocating for platform practices that do more than merely eliminate harms. The Council on Technology and Social Cohesion (techandsocialcohesion.org), for example, is a coalition that promotes a broad re-envisioning of tech structure and governance that includes a heavy emphasis on Prosocial Design (**cite**). And while we see relatively few research initiatives in general related to Prosocial Design, an exception are multiple efforts to test and promote prosocial ranking algorithms that optimize for user well being and social cohesion (e.g., Moehring et al., 2025). Working in parallel with Prosocial Design efforts is a movement of theorists and technologists who are building a "pluriverse" of interoperable platforms and "Middleware" tools centered on user agency and which provide an opening for the adoption of prosocial design features. A recent prime example of Prosocial Design meeting Middleware is the Bluesky add-on CLR:SKY, which gives users in-time suggestions to make their comments more constructive.

If Prosocial Design is gaining traction, we can only conjecture why that might be the case. It could be the political moment we are in, as Alice Hunsberger (2025) suggests, in which there is resistance toward moderation processes that are seen to inhibit free speech. Or it may be that platforms are beginning to see the limits of reactive moderation which, as Ravi Iyer (2022) has argued, sets a minimum bar for healthy engagement as the mere absence of harmful engagement.

As interest and adoption of Prosocial Design builds, we are hopeful that practice will be informed by research. Well intentioned prosocial designs do not always have their intended prosocial outcomes; in some cases they may even backfire and produce harmful outcomes (Dörr et al., 2025). To the extent possible, platforms should draw on evidence-based designs and conduct internal tests to be sure a design is, indeed, prosocial. While Trust and Safety teams are often not in a strong position to influence design decisions, anecdotally we have been told that presenting the evidence that a prosocial design is effective in producing healthier engagement can itself be an effective way to advocate for its adoption.

Equally important, public research should be informed by practice. A common observation we hear from practitioners who are interested in Prosocial Design is that much of the non-industry research they come across is of low relevance and inactionable, either because it does not address their challenges or it is unaware of platform systems and constraints. We recognize there are considerable hurdles to connecting academic researchers and practitioners, in large part because their incentives and priorities can be weakly aligned. But Trust and Safety professionals–along with other prosocial-minded technologists–can play an important role by communicating to academics and other non-industry researchers what research is needed and would be actionable. We likewise encourage academics to proactively foster connections with practitioners who can informally advise in ways to make their research more practically relevant. Such advice may be as simple as, as we have heard suggested multiple times, to include a measure of engagement in studies which can demonstrate that a prosocial design does not reduce engagement.

Finally, much more can be done to accelerate the production of publicly accessible research. Oftentimes collaborative field experiments conducted on platforms are recommended as a route to produce more quality research. Although these types of collaborations can produce the strongest evidence of the effectiveness (or ineffectiveness) of a prosocial design (see e.g., Katsaros et al., 2024; Matias, 2019), we know they present many challenges, most of which are related to building and sustaining trust: trust that a platform's reputation will not be unduly harmed, trust that researchers will have enough access to publish rigorous research, and trust that a platform's community of users will not be unethically "experimented on" without consent (e.g., Grüning et al., 2024).

As an alternate route to producing and possibly scaling high quality research, we see promise in recent innovations to foster independent research methods for testing these prosocial design interventions (Grüning et al., 2025). Such approaches include testing interventions through an external tool, such as a smartphone application (e.g., *one sec*, Grüning et al., 2023; one-sec.app), or by leveraging browser extensions that can externally modify users' platform experiences (Beknazar-Yuzbashev et al., 2025). One particularly promising route is to build simulated social media platforms explicitly designed to conduct experiments with consenting, and compensated, participants, like the *Truman*-platform (socialmedialab.cornell.edu/the-truman-platform; see e.g., Masur et al., 2021). The advent of Large Language Models increases the potential for these experimental platforms to mimic authentic social media experiences and so produce ecologically valid research (Törnberg et al., 2023; van Loon et al., 2024). These platforms, however, are expensive to build and maintain, as is paying subjects to participate in experiments. As an ambitious solution, platforms interested in discovering ways to foster healthier interactions on their platforms could support independent prosocial design research both by helping to foster these alternative methods and by funding research.

**References**

Argyle, L. P., Bail, C. A., Busby, E. C., Gubler, J. R., Howe, T., Rytting, C., Sorensen, T., & Wingate, D. (2023). Leveraging AI for democratic discourse: Chat interventions can improve online political conversations at scale. *Proceedings of the National Academy of Sciences*, *120*(41), e2311627120. https://doi.org/10.1073/pnas.2311627120

Argyle, L. P., Busby, E., Gubler, J., Bail, C., Howe, T., Rytting, C., & Wingate, D. (2023). AI chat assistants can improve conversations about divisive topics. *arXiv*. https://doi.org/10.48550/arXiv.2302.07268

Arnold, J. R., Reckendorf, A., & Wintersieck, A. L. (2021). Source alerts can reduce the harms of foreign disinformation. *Harvard Kennedy School Misinformation Review*. https://misinforeview.hks.harvard.edu/article/source-alerts-can-reduce-the-harms-of-foreign-disinformation/

Aslett, K., Guess, A. M., Bonneau, R., Nagler, J., & Tucker, J. A. (2022). News credibility labels have limited average effects on news diet quality and fail to reduce misperceptions. *Science Advances*, *8*(18), eabl3844. https://doi.org/10.1126/sciadv.abl3844

Badiei, F., & Fidler, B. (2021). The Would-Be Technocracy: Evaluating Efforts to Direct and Control Social Change with Internet Protocol Design. *Journal of Information Policy*, *11*, 376-402.

Beknazar-Yuzbashev, G., Jiménez-Durán, R., McCrosky, J., & Stalinski, M. (2025). Toxic content and user engagement on social media: Evidence from a field experiment. CESifo Working Papers, 11644. https://econpapers.repec.org/RePEc:ces:ceswps:_11644

Bor, A., Osmundsen, M., Rasmussen, S. H. R., Bechmann, A., & Petersen, M. (2020). "Fact-checking" videos reduce belief in misinformation and improve the quality of news shared on Twitter. *PsyArXiv.* https://doi.org/10.31234/osf.io/a7huq

Bowles, J., Croke, K., Larreguy, H., Liu, S., & Marshall, J. (2023). Sustaining exposure to fact-checks: Misinformation discernment, media consumption, and its political implications. *American Political Science Review*, 1-24. https://doi.org/10.1017/S0003055424001394

Brignull, H., Leiser, M., Santos, C., & Doshi, K. (2025, May 27). *Deceptive patterns – user interfaces designed to trick you*. deceptive.design. Retrieved May 27, 2025, from https://www.deceptive.design/

Celadin, T., Capraro, V., Pennycook, G., & Rand, D. G. (2023). Displaying news source trustworthiness ratings reduces sharing intentions for false news posts. *Journal of Online Trust and Safety*, *1*(5). https://dx.doi.org/10.54501/jots.v1i5.100

Chan, M. P. S., Jones, C. R., Hall Jamieson, K., & Albarracín, D. (2017). Debunking: A meta-analysis of the psychological efficacy of messages countering misinformation. *Psychological science*, *28*(11), 1531-1546. https://doi.org/10.1177/0093650219854600

Chang, J. P., Schluger, C., & Danescu-Niculescu-Mizil, C. (2022). Thread with caution: Proactively helping users assess and deescalate tension in their online discussions. *Proceedings of the ACM on human-computer interaction*, *6*(CSCW2), 1-37. https://doi.org/10.1145/3555603

Chuai, Y., Pilarski, M., Lenzini, G., & Pröllochs, N. (2024). Community notes reduce the spread of misleading posts on X. *OSF Preprints*. https://doi.org/10.31219/osf.io/3a4fe

Chuai, Y., Tian, H., Pröllochs, N., & Lenzini, G. (2023). The roll-out of community notes did not reduce engagement with misinformation on Twitter. *arXiv*. https://doi.org/10.48550/arXiv.2307.07960

Clayton, K., Blair, S., Busam, J. A., Forstner, S., Glance, J., Green, G., ... & Nyhan, B. (2020). Real solutions for fake news? Measuring the effectiveness of general warnings and fact-check tags in reducing belief in false stories on social media. *Political behavior*, *42*, 1073-1095. https://doi.org/10.1007/s11109-019-09533-0

Dörr, T., Nagpal, T., Watts, D., & Bail, C. (2025). A research agenda for encouraging prosocial behaviour on social media. *Nature Human Behaviour, 9*, 441–449. https://doi.org/10.1038/s41562-025-02102-y

Fitz, N., Kushlev, K., Jagannathan, R., Lewis, T., Paliwal, D., & Ariely, D. (2019). Batching smartphone notifications can improve well-being. *Computers in Human Behavior*, *101*, 84-94. https://doi.org/10.1016/j.chb.2019.07.016

Grüning, D. J., Kamin, J., Panizza, F., Katsaros, M., & Lorenz-Spreen, P. (2024). A framework for promoting online prosocial behavior via digital interventions. *Communications Psychology*, *2*(1), 6. https://doi.org/10.1038/s44271-023-00052-7

Grüning, D. J., Kamin, J., Saltz, E., Acosta, T., DiFranzo, D., Goldberg, B., Leavitt, A., Menczer, F., Musgrave, T., Wang, Y., & Wojcieszak, M. (2025). Independently testing prosocial interventions: Methods and recommendations from 31 researchers. *Annals of the New York Academy of Sciences*. https://doi.org/10.31234/osf.io/wvfjq_v2

Grüning, D. J., Riedel, F., & Lorenz-Spreen, P. (2023). Directing smartphone use through the self-nudge app one sec. *Proceedings of the National Academy of Sciences*, *120*(8), e2213114120. https://doi.org/10.1073/pnas.2213114120

Haliburton, L., Grüning, D. J., Riedel, F., Schmidt, A., & Terzimehić, N. (2024, May). A Longitudinal In-the-Wild Investigation of Design Frictions to Prevent Smartphone Overuse. In F. F. Mueller, P. Kyburz, J. R. Williamson, C. Sas, M. L. Wilson, P. T. Dugas, I. Shklovski (Eds.), *Proceedings of the 2024 CHI Conference on Human Factors in Computing Systems* (pp. 1-16). https://doi.org/10.1145/3613904.3642370

Hameleers, M., & van der Meer, T. (2023). Striking the balance between fake and real: under what conditions can media literacy messages that warn about misinformation maintain trust in accurate information?. *Behaviour & Information Technology*, 1-13. https://doi.org/10.1080/0144929X.2023.2267700

Hertwig, R., & Grüne-Yanoff, T. (2017). Nudging and boosting: Steering or empowering good decisions. *Perspectives on Psychological Science*, *12*(6), 973-986. https://doi.org/10.1177/1745691617702496

Herzog, S. M., & Hertwig, R. (2025). Boosting: Empowering citizens with behavioral science. *Annual Review of Psychology*, *76*, 851-881. https://doi.org/10.1146/annurev-psych-020924-124753

Hoes, E., Aitken, B., Zhang, J., Gackowski, T., & Wojcieszak, M. (2024). Prominent misinformation interventions reduce misperceptions but increase scepticism. *Nature Human Behaviour*, *8*(8), 1545-1553. https://doi.org/10.1038/s41562-024-01884-x

Horta Ribeiro, M., Cheng, J., & West, R. (2023). Automated content moderation increases adherence to community guidelines. In *Proceedings of the ACM web conference 2023* (pp. 2666-2676). https://doi.org/10.1145/3543507.3583275

Horta Ribeiro, M., West, R., Lewis, R., & Kairam, S. (2025). Post Guidance for Online Communities. *Proceedings of the ACM on Human-Computer Interaction*, *9*(2), CSCW148. https://doi.org/10.1145/3711046

Hunsberger, A. (2024). Trust & Safety is how platforms put values into action. *Everything in Moderation.* Retrieved (May 29, 2025) from https://www.everythinginmoderation.co/trust-safety-values-action/

Hunsberger, A. (2025). Is it prosocial design's time to shine?. Everything in Moderation. Retrieved (May 29, 2025) from https://www.everythinginmoderation.co/prosocial-design/

Iyer, R. (2022). *Content moderation is a dead end.* Designing Tomorrow. Retrieved (May 29, 2025) from https://psychoftech.substack.com/p/content-moderation-is-a-dead-end

Lin, H., Garro, H., Wernerfelt, N., Shore, J. C., Hughes, A., Deisenroth, D., Barr, N., Berinsky, A. J., Eckles, D., Pennycook, G., & Rand, D. G. (2024). Reducing misinformation sharing at scale using digital accuracy prompt ads. *PsyArXiv.* https://doi.org/10.31234/osf.io/u8anb

Jhaver, S., Bruckman, A., & Gilbert, E. (2019). Does transparency in moderation really matter? User behavior after content removal explanations on reddit. In *Proceedings of the ACM on Human-Computer Interaction*, 3(CSCW), 1-27. https://doi.org/10.1145/3359252

Katsaros, M., Grüning, D., & Lou, S. (2024). Offensive Comment Filtering Impact on Online Engagement: A Large-Scale Randomized Controlled Trial on Nextdoor. *PsyArXiv.* https://doi.org/10.31234/osf.io/nxuqy

Katsaros, M., Grüning, D. (in preparation). Impact of Filtering Harmful Posts on Online Engagement - A Large-Scale Randomized Controlled Trial on Nextdoor.

Katsaros, M., Yang, K., & Fratamico, L. (2022). Reconsidering tweets: Intervening during tweet creation decreases offensive content. In *Proceedings of the International AAAI Conference on Web and Social Media* (Vol. 16, pp. 477-487). https://doi.org/10.1609/icwsm.v16i1.19308

Kim, J., McDonald, C., Meosky, P., Katsaros, M., & Tyler, T. (2022). Promoting online civility through platform architecture. *Journal of Online Trust and Safety*, *1*(4). https://doi.org/10.54501/jots.v1i4.54

Lewandowsky, S., Ecker, U. K., Seifert, C. M., Schwarz, N., & Cook, J. (2012). Misinformation and its correction: Continued influence and successful debiasing. *Psychological science in the public interest*, *13*(3), 106-131. https://doi.org/10.1177/1529100612451018

Lukoff, K., Lyngs, U., Shirokova, K., Rao, R., Tian, L., Zade, H., ... & Hiniker, A. (2023). SwitchTube: A Proof-of-Concept System Introducing "Adaptable Commitment Interfaces" as a Tool for Digital Wellbeing. In *Proceedings of the 2023 CHI Conference on Human Factors in Computing Systems* (pp. 1-22). https://doi.org/10.1145/3544548.3580703

Masur, P. K., DiFranzo, D., & Bazarova, N. N. (2021). Behavioral contagion on social media: Effects of social norms, design interventions, and critical media literacy on self-disclosure. *Plos one*, *16*(7), e0254670. https://doi.org/10.1371/journal.pone.0254670

Matias, J. N. (2019). Preventing harassment and increasing group participation through social norms in 2,190 online science discussions. *Proceedings of the National Academy of Sciences*, *116*(20), 9785-9789. https://doi.org/10.1073/pnas.1813486116

Matias, J. N., Al-Kashif, R., Kamin, J., Klein, M., & Pennington, E. (2020). Volunteers Thanked Thousands of Wikipedia Editors to Learn the Effects of Receiving Thanks. *CATLab*. https://citizensandtech.org/2020/06/effects-of-saying-thanks-on-wikipedia/

Mena, P. (2020). Cleaning up social media: The effect of warning labels on likelihood of sharing false news on Facebook. *Policy & internet*, *12*(2), 165-183. https://doi.org/10.1002/poi3.214

Moehring, A., Sigerson, L., Cooper, A., Thorburn, L., Narayanan, A., Motyl, M., Ovadya, A., Eslami, M., Redmiles, E., Johnson, N. F., Allen, J., Lubin, N., Stray, J., Iyer, R., Kamin, J., Arnao, Z. (2025). *Better Feeds: Algorithms That Put People First, A How-To Guide for Platforms and Policymakers*. Knight Georgetown Institute. Retrieved (May 29, 2025) from https://kgi.georgetown.edu/wp-content/uploads/2025/02/Better-Feeds_-Algorithms-That-Put-People-First.pdf

New_ Public (n.d.) Purpose | New Public. https://newpublic.org/purpose/core-beliefs

New_Public (2022, March). *The Signals: The qualities of flourishing digital spaces.* New_Public. Retrieved (May 28, 2025) from https://docs.google.com/presentation/d/1UAsy8ZlCoRwgwOLQNblvuV5gVdPBhS2SmZVtXWT20L4/edit?slide=id.g9c2b1f0ede_1_10#slide=id.g9c2b1f0ede_1_10

Nyhan, B., & Reifler, J. (2010). When corrections fail: The persistence of political misperceptions. *Political Behavior*, *32*(2), 303-330.

Offer-Westort, M., Rosenzweig, L. R., & Athey, S. (2024). Battling the coronavirus 'infodemic' among social media users in Kenya and Nigeria. *Nature Human Behaviour*, *8*(5), 823-834. https://doi.org/10.1038/s41562-023-01810-7

Oliveira, R., Arriaga, P., Santos, F. P., Mascarenhas, S., & Paiva, A. (2021). Towards prosocial design: A scoping review of the use of robots and virtual agents to trigger prosocial behaviour. *Computers in Human Behavior*, *114*, 106547.

Papakyriakopoulos, O., Engelmann, S., & Winecoff, A. (2023, April). Upvotes? Downvotes? No Votes? Understanding the relationship between reaction mechanisms and political discourse on Reddit. In Schmidt, A., Väänänen, K., Goyal, T., Kristensson, P. O., Peters, A., Mueller, S., Williamson, J. R., Wilson, M. L. (Eds.), *Proceedings of the 2023 CHI Conference on Human Factors in Computing Systems* (pp. 1-28). https://doi.org/10.1145/3544548.3580644

Pennycook, G., Bear, A., Collins, E. T., & Rand, D. G. (2020). The implied truth effect: Attaching warnings to a subset of fake news headlines increases perceived accuracy of headlines without warnings. *Management science*, *66*(11), 4944-4957. https://doi.org/10.1287/mnsc.2019.3478

Pennycook, G., & Rand, D.G. (2022). Accuracy prompts are a replicable and generalizable approach for reducing the spread of misinformation. *Nature Communications, 13*, 2333. https://doi.org/10.1038/s41467-022-30073-5

Pereira, F. B., Bueno, N. S., Nunes, F., & Pavão, N. (2024). Inoculation reduces misinformation: experimental evidence from multidimensional interventions in brazil. *Journal of Experimental Political Science*, *11*(3), 239-250. https://doi.org/10.1017/XPS.2023.11

Porter, E., & Wood, T. J. (2022). Political misinformation and factual corrections on the Facebook news feed: Experimental evidence. *The Journal of Politics*, *84*(3), 1812-1817. https://doi.org/10.1086/719271

Prosocial Design Network (2025). Digital Intervention Library. Prosocial Design Network [Digital resource]. https://doi.org/10.17605/OSF.IO/Q4RMB

Randazzo, C., Scott, C. F., Bellini, R., Ammari, T., Devito, M. A., Semaan, B., & Andalibi, N. (2023). Trauma-Informed Design: A Collaborative Approach to Building Safer Online Spaces. In C. Fiesler, L. Terveen, M. Ames, S. Fussell, E. Gilbert, V. Liao, X. Ma, X. Page, M. Rouncefield, V. Singh, P. Wisneski (Eds.), *Companion Publication of the 2023 Conference on Computer Supported Cooperative Work and Social Computing* (pp. 470-475). https://doi.org/10.1145/3584931.3611277

Renault, T., Amariles, D. R., & Troussel, A. (2024). Collaboratively adding context to social media posts reduces the sharing of false news. *arXiv.* https://doi.org/10.48550/arXiv.2404.02803

Reviligio, U., & Giovanardi, M. (2025). Advancing 'prosocial tech design' and shaping the EU's platform design governance. Robert Schuman Centre. Retrieved (May 29, 2025) from https://techandsocialcohesion.org/wp-content/uploads/2025/04/EUI-Policy-Brief-Advancing-Prosocial-Tech-Design-and-shaping-the-EU-platform-design-governance.pdf

Ribeiro, M. H., West, R., Lewis, R., & Kairam, S. (2024). Post Guidance for Online Communities. *arXiv.* https://arxiv.org/abs/2411.16814

Roozenbeek, J., & Van der Linden, S. (2019). Fake news game confers psychological resistance against online misinformation. *Palgrave Communications*, *5*(1), 1-10. https://doi.org/10.1057/s41599-019-0279-9

Roozenbeek, J., Van Der Linden, S., Goldberg, B., Rathje, S., & Lewandowsky, S. (2022). Psychological inoculation improves resilience against misinformation on social media. *Science advances*, *8*(34), eabo6254. https://doi.org/10.1126/sciadv.abo6254

Schirch, L., Iyer ,R., Slachmuijlder, L. (2023). Toward Prosocial Tech Design Governance, Tech Policy Press. https://www.techpolicy.press/toward-prosocial-tech-design-governance/

Schirch, L. (2025). Blueprint on Prosocial Tech Design Governance. Council on Technology and Social Cohesion with University of Notre Dame and Toda Peace Institute. May 2025.

Schmidt, A. T., & Engelen, B. (2020). The ethics of nudging: An overview. *Philosophy compass*, *15*(4), e12658.

Simon, G. (2020). OpenWeb tests the impact of "nudges" in online discussions. *OpenWeb Blog*. https://www.openweb.com/blog/openweb-improves-community-health-with-real-time-feedback-powered-by-jigsaws-perspective-api

Srinivasan, K. B., Danescu-Niculescu-Mizil, C., Lee, L., & Tan, C. (2019). Content removal as a moderation strategy: Compliance and other outcomes in the changemyview community. *Proceedings of the ACM on Human-Computer Interaction*, *3*(CSCW), 1-21. https://doi.org/10.1145/3359265

Stroud, N. J., Muddiman, A., & Scacco, J. M. (2017). Like, recommend, or respect? Altering political behavior in news comment sections. *New media & society*, *19*(11), 1727-1743. https://doi.org/10.1177/1461444816642420

Törnberg, P., Valeeva, D., Uitermark, J., & Bail, C. (2023). Simulating social media using large language models to evaluate alternative news feed algorithms. *arXiv.* https://doi.org/10.48550/arXiv.2310.05984

Trust and Safety Professional Association. "Trust and Safety Curriculum." Accessed [May 29, 2025]. https://www.tspa.org/curriculum/ts-curriculum/

Tyler, T. R. (2016). Procedural justice. In D. A. Krauss (Ed.), *Jury Psychology: Social Aspects of Trial Processes* (pp. 25-40). https://doi.org/10.4324/9781315590790

Tyler, T., Katsaros, M., Meares, T., & Venkatesh, S. (2021). Social media governance: can social media companies motivate voluntary rule following behavior among their users?. *Journal of experimental criminology*, *17*, 109-127. https://doi.org/10.1007/s11292-019-09392-z

van Loon, A., Katta, S., Bail, C. A., Hillygus, D. S., & Volfovsky, A. (2024). Designing social media to promote productive political dialogue on a new research platform. *PsyArXiv.* https://doi.org/10.31235/osf.io/dngcj

Vriens, E., & Andrighetto, G. (2024). Why social norms may fail us when we need them most. *Current Opinion in Psychology, 62*, 101975. https://doi.org/10.1016/j.copsyc.2024.101975

Walter, N., & Tukachinsky, R. (2020). A meta-analytic examination of the continued influence of misinformation in the face of correction: How powerful is it, why does it happen, and how to stop it?. *Communication Research*, *47*(2), 155-177. https://doi.org/10.1177/0093650219854600

Wang, Y., & Diakopoulos, N. (2022). Highlighting high-quality content as a moderation strategy: The role of new york times picks in comment quality and engagement. *ACM Transactions on Social Computing (TSC)*, *4*(4), 1-24. https://doi.org/10.1145/3484245

Weijun C., Yan Q., Yuwen Z., Christina B., Akos L., Harivardan J. (2022). Notifications: why less is more — how Facebook has been increasing both user satisfaction and app usage by sending only a few notifications. *Medium.* https://medium.com/@AnalyticsAtMeta/notifications-why-less-is-more-how-facebook-has-been-increasing-both-user-satisfaction-and-app-9463f7325e7d

Wojcik, S., Hilgard, S., Judd, N., Mocanu, D., Ragain, S., Hunzaker, M. B., ... & Baxter, J. (2022). Birdwatch: Crowd wisdom and bridging algorithms can inform understanding and reduce the spread of misinformation. *arXiv.* https://doi.org/10.48550/arXiv.2210.15723

Yadav, M., & Garg, N. Learning to resist misinformation: A field experiment. *Beyond Moderation: Emerging Research in Online Governance*, 19-25. https://static1.squarespace.com/static/60a67985a6f31b7cb71aeab0/t/64dd40d89f5ac067ce616861/1692221657255/SMGI+Beyond+Moderation+Essays.pdf#page=19